\documentclass[aps,twocolumn,pra,superscriptaddress,amsmath,showpacs,tightenlines]{revtex4-1}
\usepackage{amssymb}
\usepackage{amsmath}
\usepackage{graphicx}
\usepackage{subfigure}
\usepackage{natbib}
\usepackage{epsfig}
\usepackage{amsfonts}
\usepackage{mathrsfs}
\usepackage{xcolor}
\usepackage[toc,page,title,titletoc,header]{appendix}

\begin{document}
\title{Superconducting giant atom waveguide QED: Quantum Zeno and Anti-Zeno effects in ultrastrong coupling regime}
\author{Xiaojun Zhang}
\affiliation{Center for Quantum Sciences and School of Physics, Northeast Normal University,
Changchun 130024, China}
\author{Weijun Cheng}
\affiliation{Center for Quantum Sciences and School of Physics, Northeast Normal University,
Changchun 130024, China}
\author{Zhirui Gong}
\affiliation{College of Physics and Optoelectronic Engineering, Shenzhen University, Shenzhen, 518060, China}
\author{Taiyu  Zheng}
\email{zhengty@nenu.edu.cn}
\affiliation{Center for Quantum Sciences and School of Physics, Northeast Normal University,
Changchun 130024, China}
\author{Zhihai Wang}
\email{wangzh761@nenu.edu.cn}
\affiliation{Center for Quantum Sciences and School of Physics, Northeast Normal University,
Changchun 130024, China}

\begin{abstract}
The giant atom system is a new paradigm in quantum optics, in which the traditional dipole approximation is not available. In this paper, we construct an artificial giant atom model by coupling a superconducting circuits to a transmission line by two coupling points. In the ultrastrong coupling regime, we show that the Lamb shift of the giant atom, which is induced by the non-negligible counter-rotating  atom-waveguide coupling term, will modify its dissipation process. Furthermore, we investigate quantum Zeno and anti-Zeno effect where the size of the giant atom serves as a sensitive controller. Specifically, by comparing the critical measurement interval and the life time of the giant atom, we clarify the condition for the occurring of quantum anti-Zeno effect. We hope our work is useful for the application of giant atom system in the investigation of fundamental problems of quantum mechanics.

\end{abstract}
%\pacs{42.50.Pq, 03.67.Lx, 42.50.Dv}
\maketitle
\section{introduction}
The waveguide QED with artificial giant atom is a recently developed topics in the field of quantum optics~\cite{review,Nat14}. In the giant atom setup, the artificial atom interacts with the waveguide non-locally, that is, via two or more coupling points. This nonlocal coupling leads to lots of otherwise nonexist phenomenon, such as frequency dependent relaxation and Lamb shift~\cite{Lambpra14}, non-Markovian atomic dissipation\cite{LG2017,GA2019,Guoprr20,SG2020}, exotic charity~\cite{XW2021,AS2021} as well as the decoherence free interatomic
interaction~\cite{AF2018,AC2020}. Besides, due to the interference effect, the photon propagation behavior can also be modified by the location or the size of the giant atom~\cite{WZ,Jia1,Jia2,LD1,LD2,Liao,Yuan}.

The giant atom setup is first experimentally realized by nonlocally coupling a transmon qubit to the surface acoustic wave via interdigitated transducer~\cite{Nat14}. Due to the slow velocity of the acoustic wave, the size of the transmon is ten times of the wavelength, to form a real giant atom. On the other hand, the superconducting qubit can also be nonlocally coupled to the microwave transmission line via wire, capacitance or inductance~\cite{BK2020,AM2021}. Although the qubit is smaller than the wavelength of the microwave, the distance of the coupling points can be arbitrary large, so that we can still reach a giant atom setup.

The superconducting qubit (or artificial atom) is a promised candidate to realize quantum information processing thanks to its long coherence time and easier integration~\cite{super1,super2,super3,super4,super5,super6,super7}. Moreover, in the circuit QED setup, the ultrastrong and deep strong coupling between the superconducting qubit and resonator has been proposed and realized where the light-matter coupling strength can be comparable to or even surpass their frequencies~\cite{ultra1,ultra2,ultra3,ultra4,ultra5}. In this case, the counter rotating wave term can not be neglected in the interaction Hamiltonian and it will lead some exotic quantum phenomenon, such as Bloch-Siegert shift~\cite{TN2010,PF2010}, ground state degeneration,quantum phase transition~\cite{PN2010,MJ2015,MJ2016,ML2017} and quantum vacuum emission~\cite{SD2009}, just to name a few.

Based on the above achievements, we here discuss the exotic quantum effect in a superconducting
giant atom setup. Specifically, we investigate the quantum Zeno effect (QZE) and anti-Zeno effect (QAZE)~\cite{zeno1,zeno2,zeno3,zeno4,zeno5,zeno6,zeno7,zeno8,zeno9,zeno10} in a waveguide QED with superconducting circuits which acts as an effective two-level giant atom and couples to the transmission line waveguide via two connecting points. In the ultrastrong coupling regime, we find that the counter rotating wave term will induce a Lamb shift for the giant atom. Therefore, the atomic emission behavior induced by the waveguide is modified compared with that within rotating wave approximation~\cite{HY2021}. Furthermore, we introduce the frequent measurement on the system, and find the QZE (QAZE), which implies that the measurement will decrease (increase) the dissipation of the giant atom system. We clarify that the ratio between the measurement interval and the decay time of the atom is a key element in the judgement of QAZE. In this manner, we find the parameter regime in which the QAZE happens for different atomic size.

The rest of the paper is organized as follows. In Sec.~\ref{model}, we present our model and derive the effective Hamiltonian. In Sec.~\ref{waveguide}, we obtain the effective Hamiltonian
by use of transformed rotating wave approximation. In Sec.~\ref{zeno}, we discuss the QZE and QAZE in our model and give a short summary in Sec.~\ref{conclusion}.

\section{Superconducting circuits}
\label{model}
 \begin{figure}
  \centering
\includegraphics[width=8cm]{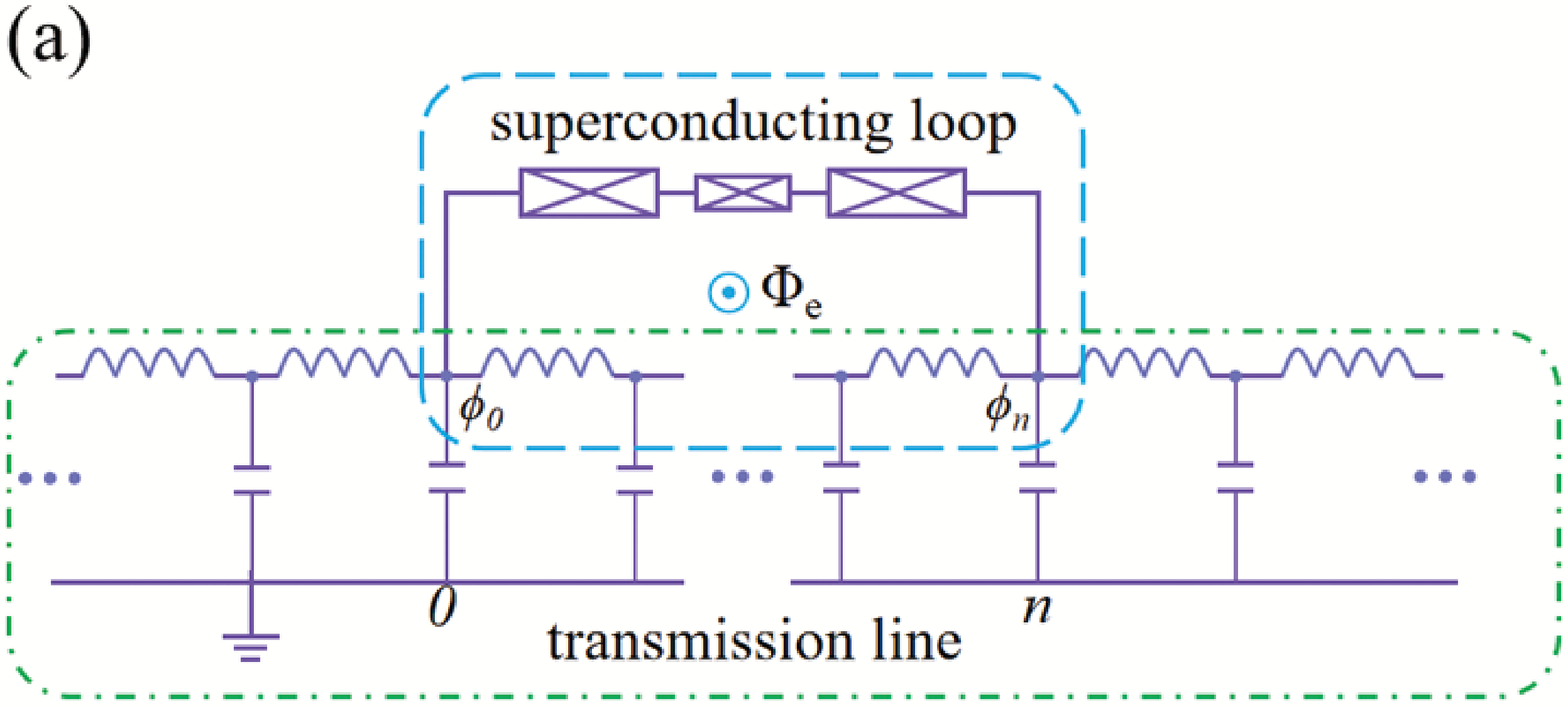}\nonumber\\
\includegraphics[width=8cm]{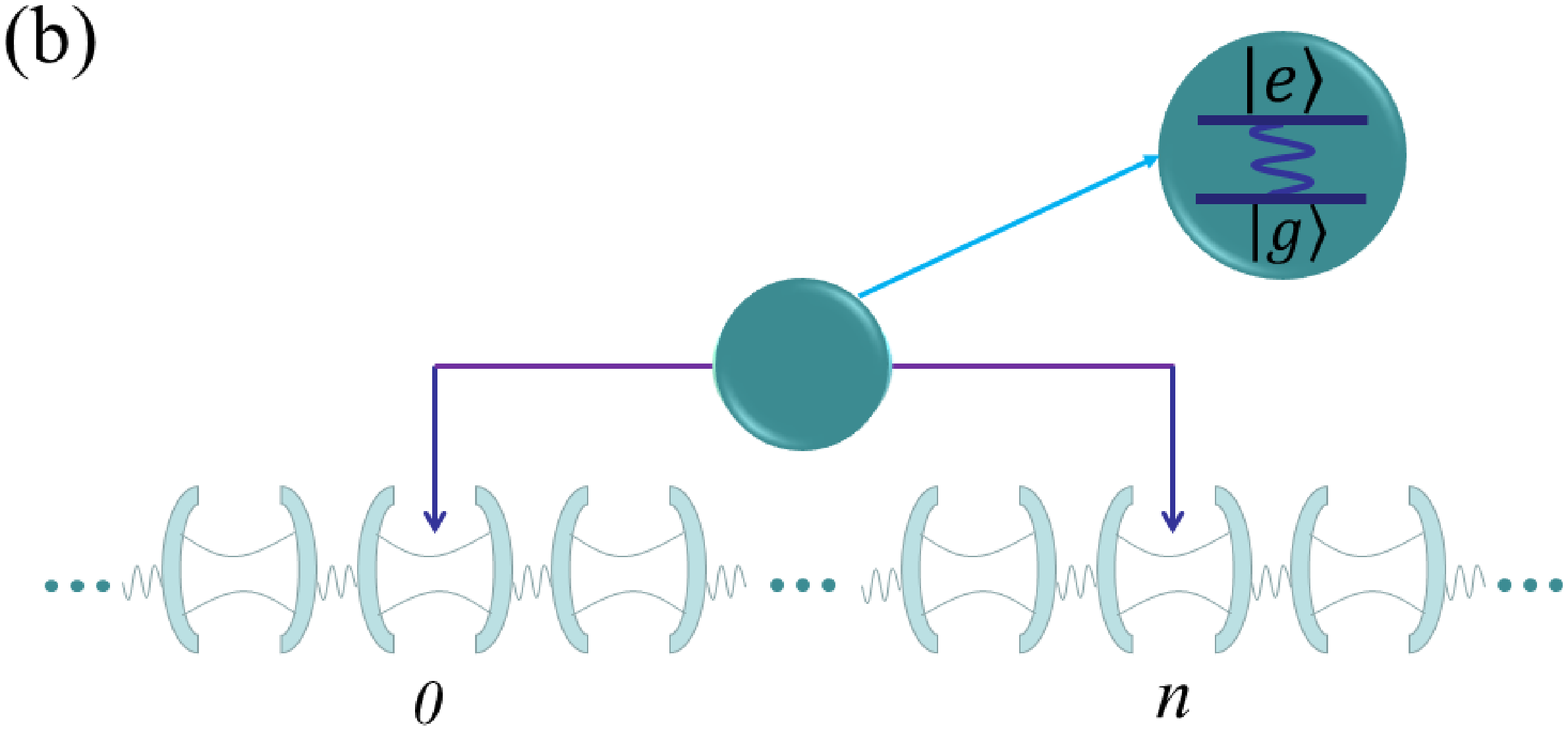}\nonumber\\
\caption{The setup of waveguide QED system with a giant atom via two coupling points. (a) The coupling diagram between a superconduting circuits and transmission line. (b) The effective model for a giant atom coupling to  coupled to a photonic lattice via $0$th and $n$th sites.}
\label{scheme}
\end{figure}
As sketched in Fig.~\ref{scheme}(a), the system we consider is composed by a superconducting loop with three Josephson junctions, which couples to a transmission line.  The Lagrangian of the system is written as
$\mathcal{L}=\mathcal{L}_J+\mathcal{L}_s$, where~\cite{ultra2}
\begin{eqnarray}
\mathcal{L}_J&=& \frac{1}{2}{C_J}{(\frac{{{\Phi _0}}}{{2\pi }})^2}\left(\dot \phi _{J1}^2 +\dot \phi _{J3}^2+\alpha\dot \phi _{J2}^2\right)\nonumber\\&&+ {E_J}\left(\cos {\phi _{J1}} + \cos {\phi _{J3}}+\alpha \cos {\phi _{J2}}\right)
\end{eqnarray}
is the Lagrangian for the superconducting loop. Here, $\phi_{Ji}\,(i=1,2,3)$ is the phase differences between the two ends of the $i$th Josephson junction.  The two end Josephson junctions have
equal Josephson energies $E_J$ and capacitances $C_J$ while the middle has $\alpha E_J$ and $\alpha C_J$ ($1/2<\alpha<1$).

The second part $\mathcal{L}_s$~\cite{AA2019}
\begin{equation}
\mathcal{L}_s=\sum\limits_i {\frac{{\Delta x{C_T}}}{2}{{(\frac{{{\Phi _0}}}{{2\pi }})}^2}} \dot \phi _i^2 - \sum\limits_i {{{(\frac{{{\Phi _0}}}{{2\pi }})}^2}\frac{{{{({\phi _{i + 1}} - {\phi _i})}^2}}}{{2\Delta x{L_T}}}}
\end{equation}
describes the transmission line, where $C_T$ and $L_T$ are the capacitance and inductance per unit length respectively, {and $(\Phi_0/2\pi)\phi_i$ is the flux of the $i$th node in the transmission line as shown in Fig.~\ref{scheme} (a).}

Furthermore, we connect the superconducting loop to the transmission line via the $0$th and $N$th nodes by wires and the closed loop is threaded by an external magnetic flux $\Phi_e$.
The phases satisfy the quantization condition of the magnetic flux, that is,
\begin{equation}
\phi _{J1}+ \phi _{J2} + \phi _{J3} - \Delta \phi  = 2f\pi
\end{equation}
where $\Delta \phi=\phi_0-\phi_N$ is the flux difference between the $0$th and $N$th nodes in the transmission line.  $f=\Phi_e/\Phi_0$ is the reduced magnetic flux while $\Phi_0=\pi\hbar/e$ is the flux quanta.

Defining the conjugate momentum of the flux as $P_{Ji}=\partial {\mathcal{L}}/\partial{\dot \phi_{Ji}}$ and $P_{i}=\partial {\mathcal{L}}/\partial{\dot \phi_{i}}$, which have the unit of charge, it then yields $\vec {P}=\mathcal{M}\dot{\vec{\phi}}$ where ${\vec P}=(P_N,P_0,P_{J+})^T$,
and ${\vec \phi}=(\phi_N,\phi_0,\phi_{J+})^T$ and the $3\times3$ matrix $M$ reads
{{\begin{equation}
 \mathcal{M}=\left( {\begin{array}{*{20}{c}}
\alpha \mathcal{C}_J+ \Delta x\mathcal{C}_T&-\alpha \mathcal{C}_J&-\alpha \mathcal{C}_J\\-\alpha \mathcal{C}_J&\alpha \mathcal{C}_J+ \Delta x\mathcal{C}_T&\alpha \mathcal{C}_J\\
-\alpha \mathcal{C}_J&\alpha \mathcal{C}_J&{(\alpha+\frac{1}{2})\mathcal{C}_J}
\end{array}} \right).
\end{equation}}}

In the above equations, we have defined $\mathcal{C}_J=C_J(\Phi_0/2\pi)^2,\mathcal{C}_T=C_T(\Phi_0/2\pi)^2$ and $\phi_{J\pm}=\phi_{J1}\pm\phi_{J3}$. The Hamiltonian of the whole circuits
 \begin{equation}
 H=\sum_{i=1}^3\left(\dot \phi_{Ji} P_{Ji}\right)+\sum_i\left(\dot \phi_{i} P_{i}\right )-\mathcal{L}
 \end{equation}
 yields $H=H_s+H_t+H_I$ with
 \begin{eqnarray}
 H_s&=&\frac{P_{J+}^2}{M_{+}} + \frac{{P_ {J-} ^2}}{{{C_J}{{(\frac{{{\Phi _{0}}}}{{2\pi }})}^2}}}+{E_J}[\alpha \cos (2f\pi  - {\phi _ {J+} })\nonumber\\&&-2\cos \frac{{{\phi _ {J+} }}}{2}\cos \frac{{{\phi _ {J-} }}}{2}]+O(\Delta \phi^2),\\
 H_t&=&\sum\limits_{i} \frac{{P_i^2}}{{2\Delta x{C_T}{{(\frac{{{\Phi _0}}}{{2\pi }})}^2}}}+\sum\limits_{i}{{{(\frac{{{\Phi _0}}}{{2\pi }})}^2}\frac{{{{({\phi _{i + 1}}-{\phi _i})}^2}}}{{2\Delta x{L_T}}}}\nonumber \\&&+ J_{0n}{P_0}{P_n},\\
 H_I&=&J\left({P_0}{P_{J+}} - {P_n}{P_ {J+}}\right)+\alpha E_J\Delta\phi\sin(\phi_{J+}).
 \end{eqnarray}
Here, $H_s$, $H_t$ and $H_I$ are respectively the Hamiltonian for the superconducting loop, the transmission line and their coherent coupling. In the above equation, we have defined ($N=\mathcal{M}^{-1}$)
{\begin{eqnarray}
{M_ +} &=&\{\frac{1}{2}\mathcal{C}_J(\alpha+\frac{1}{2})N_{33}^2
+(\alpha\mathcal{C}_J+\Delta x\mathcal{C}_T)N_{13}^2\nonumber \\&&+ \alpha\mathcal{C}_J(N_{13}^2-2N_{13}N_{33})\}^{-1}\\
{J_{0n}} &=& -\mathcal{C}_J(\alpha+\frac{1}{2})N_{13}^2
+2(\alpha \mathcal{C}_J+\Delta x\mathcal{C}_T)N_{11}N_{12}\nonumber\\
&&+\alpha\mathcal{C}_J(2N_{11}N_{13}-2N_{12}N_{13}-N_{11}^{2}-N_{12}^{2}) \nonumber\\
J &=&-(\alpha+\frac{1}{2})\mathcal{C}_JN_{13}N_{33}\nonumber \\&&+(\alpha \mathcal{C}_J + \Delta x\mathcal{C}_T)(N_{12}-N_{11})N_{13}
\nonumber\\&& + \alpha \mathcal{C}_J[(
N_{12}-N_{11})(N_{13}-N_{33})+2N_{13}^2].
\end{eqnarray}}

\begin{figure}
   \includegraphics[width=8.5cm]{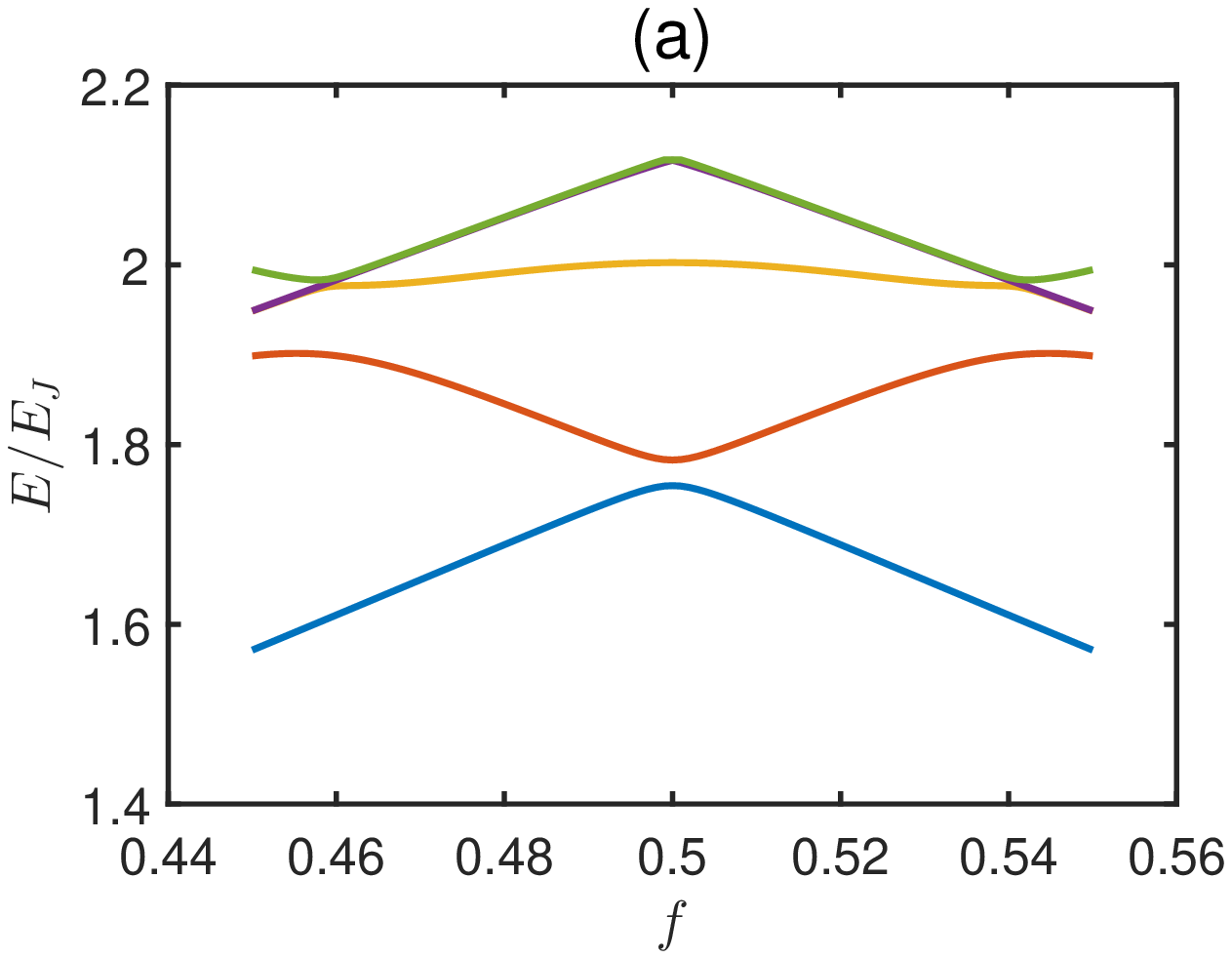}\nonumber\\
    \includegraphics[width=8.5cm]{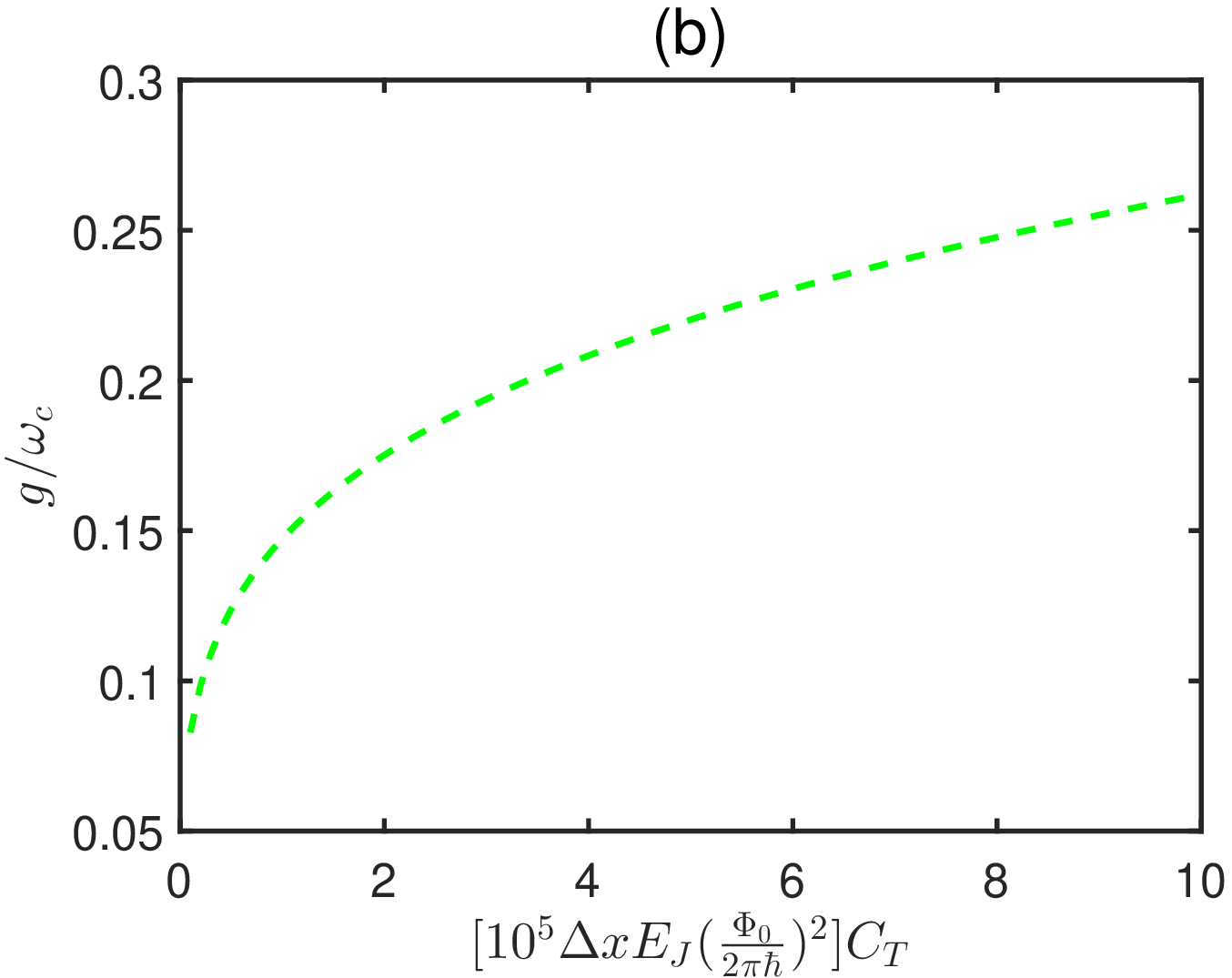}\nonumber\\
    \caption{(a) The lowest five energy levels of the Hamiltonian $H_s$. (b) The effective atom-waveguide coupling strength $g$ as a function of $C_T$. {The parameters are set as $\alpha=0.8,f=0.5.$} }
    \label{energylevel}
\end{figure}

To quantize the magnetic fluxes and their conjugate momentum in a canonical way, we promote both of the superconducting loop and the transmission line variables to operators and impose the commutation relation
$[\phi_{Ji},P_{Jj}]=i\hbar\delta_{ij},[\phi_{Ji},\phi_{Jj}]=[P_{Ji},P_{Jj}]=0$
for $i,j=+,-$ and
$[\phi_{m},P_{n}]=i\hbar\delta_{mn},[\phi_{m},\phi_{n}]=[P_{m},P_{n}]=0$.

In Fig.~2(a), we plot the lowest five levels of the Hamiltonian $H_s$ as a function of the reduced magnetic flux $f$. It shows that, when $f=0.5$, the lowest two energy levels are well separated from the higher ones. In this situation, the superconducting loop can be considered as a two-level system with the energy-level spacing {$\Delta_E=0.0288E_J$}.

The second part of the Hamiltonian $H_t$ describes the free energy and the interactions inside the transmission line. The first line, which is obtained by the Legendre transformation to the  Lagrangian $\mathcal{L}_t$, represents the bare energy of the transmission line. The second line is the superconducting loop induced effective interaction between the $0$th and $n$th nodes.

The last part $H_I$ shows that the superconducting loop couples to the transmission line via two separative ports with $i=0$ and $i=n$. The Hamiltonian shows that they couple each other at two separated nodes. In the parameter regime we consider in this paper, the coupling strengths satisfy {$J_{0n}\approx10^{-12}E_J, J\approx10^{-6}E_J$}, therefore, the last term in $H_t$ and the first term in $H_I$ can be safely neglected, and the Hamiltonian becomes
\begin{eqnarray}
H_t&\approx&\sum\limits_{i} \frac{{P_i^2}}{{2\Delta x{C_T}{{(\frac{{{\Phi _0}}}{{2\pi }})}^2}}}+\sum\limits_{i}{{{(\frac{{{\Phi _0}}}{{2\pi }})}^2}\frac{{{{({\phi _{i + 1}}-{\phi _i})}^2}}}{{2\Delta x{L_T}}}}, \\
 H_I&\approx&\alpha E_J\Delta\phi\sin(\phi_{J+}),
\end{eqnarray}
which is the beginning of the following discussions.

\section{Giant atom-waveguide QED model}
\label{waveguide}

To investigate the quantum effect in the above circuits, we note that the superconducting loop is equivalent to an artificial two-level system, and the transmission line is equivalent to a coupled resonator array waveguide. Different from the conventional waveguide QED setup with natural atom~\cite{Rabl2016}, the artificial two-level system here couples to the waveguide via two connecting points. We will name such artificial two-level system as ``giant atom'' in what follows, and
denote its excited and ground states as $|e\rangle$ and $|g\rangle$ respectively. Introducing the annihilation and creation operators
\begin{eqnarray}
a_j &=& \frac{{{\phi _j}}}{{2{{(\frac{1}{2}{{(\frac{{2\pi }}{{{\Phi _0}}})}^2}\sqrt {\frac{{{L_T}}}{{2{C_T}}}} )}^{\frac{1}{2}}}}} + \frac{{i{P_j}}}{{2{{(\frac{1}{2}{{(\frac{{{\Phi _0}}}{{2\pi }})}^2}\sqrt {\frac{{2{C_T}}}{{{L_T}}}} )}^{\frac{1}{2}}}}}, \\ a_j^{\dagger}&=&\frac{{{\phi _j}}}{{2{{(\frac{1}{2}{{(\frac{{2\pi }}{{{\Phi _0}}})}^2}\sqrt {\frac{{{L_T}}}{{2{C_T}}}} )}^{\frac{1}{2}}}}} -\frac{{i{P_j}}}{{2{{(\frac{1}{2}{{(\frac{{{\Phi _0}}}{{2\pi }})}^2}\sqrt {\frac{{2{C_T}}}{{{L_T}}}} )}^{\frac{1}{2}}}}},
\end{eqnarray}
 for the $j$th discrete circuit in the transmission line, the second quantization Hamiltonian of the equivalent waveguide QED setup is obtained as $H=H_c+H_a+H_I$, where
    \begin{eqnarray}
{H_c} &=& {\omega _c}\sum\limits_j {a_j^\dag } {a_j} - \xi \sum\limits_j {\left( {a_{j + 1}^\dag {a_j} + a_j^\dag {a_{j + 1}}} \right)}, \nonumber\\
{H_a}&=& \frac{1}{2}\Omega {\sigma _z} ,\nonumber\\
{H_I}&=& g\left[ {\left( {{a_0} + a_0^\dag } \right){\sigma _x} + \left( {{a_n} + a_n^\dag } \right){\sigma _x}} \right].
\end{eqnarray}
Herein, $\sigma_{x}=|e\rangle\langle g|+|g\rangle\langle e|$ and $\sigma_{z}=|e\rangle\langle e|-|g\rangle\langle g|$ are the Pauli matrices for the giant atom. $\Omega$ is the energy-level spacing of the lowest two energy levels in Fig.~\ref{energylevel}(a) at $f=0.5$. The bare frequency of the each resonator and the hopping strength between nearest resonators are respectively
\begin{eqnarray}
\omega_c&=&\sqrt {\frac{2}{{\Delta {x^2}{L_T}{C_T}}}},\\
\xi&=&\sqrt {\frac{1}{{8\Delta {x^2}{L_T}{C_T}}}},
\end{eqnarray}
and
\begin{equation}
g=\frac{{\alpha {E_J}}}{{{{(\frac{{{\Phi _0}}}{{2\pi }})}^2}}}\sqrt[4]{{\frac{{{L_T}}}{{8{C_T}}}}}\left\langle e \right|\sin {\phi _{J + }}\left| g \right\rangle
\end{equation}
is the atom-waveguide coupling strength. In our consideration, the inter-resonator coupling strength $\xi$ can be much smaller than the frequency $\omega_c$, and we have applied the rotating wave approximation in $H_c$. However, as shown in Fig.~\ref{energylevel}(b), where the atom-waveguide coupling strength $g$ is plotted as a function of {$C_{T}$}, $g/\omega_c$ can be larger than $0.1$ for the given  parameter regime, that is, we reach the ultrastrong coupling. As a result, we have to take the counter-rotating  wave terms in $H_I$ into consideration in the following discussions.

The counter-rotating wave terms in the atom-waveguide coupling Hamiltonian breaks the excitation conservation. To deal with this issue, we here apply the transformed rotating wave approximation, which actually eliminates the counter-rotating coupling terms in a transformed
representation by choosing an on-demand unitary transformation~\cite{Gan2010,Yu2012,Zhang2015,
Wang2018}. To this end, we begin with the
Hamiltonian in the momentum space by introducing Fourier transformation ${a_j} = \sum_k a_k e^{ikj}/\sqrt{{N_c}}$, where $N_c\rightarrow\infty$ is the length of the waveguide,  and the Hamiltonian yields $H=H_0+H_I$, where
\begin{eqnarray}
H_0&=&\sum_k \omega _k a_k^{\dag} a_k+\frac{1}{2}\Omega {\sigma _z},\\
H_I&=&\sigma_x\sum_k\left(g_k a_k^{\dag} +{\rm H.c.}\right).
\end{eqnarray}
Here, ${\omega _k} = {\omega _c} - 2\xi \cos k$ is the frequency of the $k$th mode in the waveguide and $g_k=g(1 + e^{ikn})/\sqrt{N_c}$ is its coupling strength with the giant atom.

Following the spirit of Fr\"{o}hlish-Nakajima transformation, which is also named as Schrieffer-Wolff transformation~\cite{Zhu12000,Zhu22000,Li2007,MB2009}, we perform the unitary transformation of $H_{\rm eff}= {e^S}H{e^{ - S}}$ with~\cite{AQpra10}
\begin{equation}
S = \sum\limits_k ({{A_k^{*}}a_k^\dag } {\sigma^+} - {A_k}{a_k}\sigma^{-})
\end{equation}
where {$\sigma^{\pm}=(\sigma_x\pm i\sigma_y)/2$} and $A_k$ is to be determined.
Up to the second order of $g_k$, the effective Hamiltonian is given as
  \begin{equation}
   {H_{\rm eff}}={H_0} + {H_1} + \frac{1}{2}\left[ {{H_1},S} \right] + \frac{1}{2}\left[ {{H_I},S} \right] +...
  \end{equation}
To eliminate the counter-rotating wave coupling $a_k^\dag {\sigma ^ + } + {\rm H.c.}$ in the first order term $H_1= {H_I}+\left[ {{H_0},S} \right]$, we need
\begin{eqnarray}
A_k &=&  - \frac{{{g_k}}}{{{\omega _k} + \Omega }}.
\end{eqnarray}
As a result, the effective Hamiltonian becomes
  \begin{eqnarray}
{H_{\rm eff}} &\approx& \sum\limits_k {{\omega _k}a_k^\dag } {a_k}\nonumber + \frac{{{\Omega _1}}}{2}{\sigma _z} + \sum\limits_k {\left( {{g_k}{a_k^{\dagger}}{\sigma^-} + {\rm H.c.} } \right)},
\label{eff}\\
  \end{eqnarray}
where we have omitted the intercrossing terms such as $a_k^\dag a_{k'}^\dag$ and $a_k a_{k'}$ due to their high frequencies (in the view of interaction presentation). $\Omega_1$ is the modified level spacing for the giant atom satisfying $\Omega_1=\Omega+\delta$, where $\delta$ is
ultrastrong coupling induced Lamb shift with
  \begin{eqnarray}
\delta&=&\sum\limits_k \frac{{{{\left| {{g_k}} \right|}^2}}}{{{\omega _k} + \Omega }}\left(1+2a_k^{\dagger}a_k\right)\nonumber\\
&\approx&\sum\limits_k \frac{{{{\left| {{g_k}} \right|}^2}}}{{{\omega _k} + \Omega}}\nonumber \\
&=&\frac{g^2}{\pi}\int_{-\pi}^{\pi}dk\frac{1+\cos(kn)}{\omega+\Omega-2\xi\cos k}.\label{shifte}
  \end{eqnarray}

\begin{figure}
  \centering
  \includegraphics[width=8.5cm]{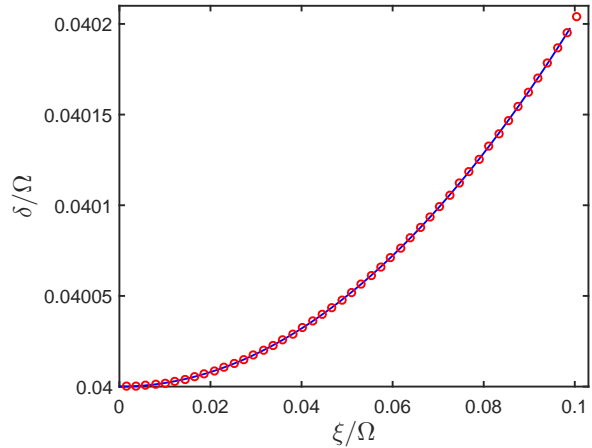}\\
  \caption{Lamb shift $\delta$ as a function $\xi$. The parameters are set as $\omega_{c}=\Omega$, $g=0.2\Omega$, $n=4$. The solid line comes from Eq.~(\ref{shifte}) and the empty circle is the fitting result.}\label{shift}
\end{figure}

We consider the situation that the atom is initially excited and all of the resonators are in their vacuum state, the term $a_l^{\dagger}a_l$ only contributes a small modification to the $l$th mode shift. It is obvious that the shift is proportional to $g^2$, which is subject to a second order perturbation effect. The numerical integration shows that the shift $\delta$ is nearly independent of the atomic size when $n>2$. Furthermore, we plot the energy shift as a function of $\xi$ in Fig.~\ref{shift} for $n=4$. It shows that the shift is nearly a quadratic function of $\xi$ and a numerical fitting yields {$\delta\approx 0.02\xi^2+0.04$, which is given by the empty circles in Fig.~\ref{shift}. It shows that the shift is $\delta\approx0.4\xi$ even in the ultrastrong coupling regime $g=0.2\Omega$. Remember that, the energy band of the waveguide is centered at $\omega$ with a width $4\xi$, so the modified atomic frequency $\Omega_1$ is still inside the energy band of the waveguide and far way from the edge of the band. Therefore, considering the waveguide as a structured environment in what follows, we can still apply the Markovian approximation based on the effective atom-waveguide interaction   Hamiltonian in Eq.~(\ref{eff}).}

\section{Quantum Zeno and anti-Zeno effect}
\label{zeno}

The QZE (QAZE) tells us that frequent measurements will slow down (speed up) the decay of a system.  In this paper, we will investigate the QZE and QAZE in our system both beyond and within the rotating wave approximation (RWA)~\cite{AQpra10,ZHprl08,ZHpra10,SHpra19,SHpra17}.

Let us begin beyond the RWA  based on the effective Hamiltonian in
Eq.~(\ref{eff}). We choose the initial state as $|\psi(0)\rangle=|e,G\rangle$ which represents that the giant atom is initially excited and all of the resonators in the waveguide are in the vacuum states. The initial state is not changed by the unitary transformation $\exp(-S)$, that is,
\begin{equation}
{e^{ - S}}\left| {e,G} \right\rangle  = \left( {1 - S + \frac{1}{2}{S^2}} \right)\left| {e,G} \right\rangle  = \left| {e,G} \right\rangle.
  \end{equation}
During the time evolution which is governed by $H_{\rm eff}$, the atomic survival probability $P_e(t)$ is formally expressed as
\begin{equation}
P_e(t)={\rm Tr}\left(|e\rangle\langle e|e^{-iH_{\rm eff}t}|e,G\rangle\langle e,G|e^{iH_{\rm eff}t}\right),
\end{equation}
Neglecting the transformation induced multiphoton process~\cite{AQpra10}, the above equation is approximated  as
\begin{eqnarray}
P_e(t)&\approx&|\langle e,G|e^S e^{-iH_{\rm eff}t}e^{-S}|e,G\rangle|^2\nonumber \\
&=&|\langle e,G|e^{-iH_{\rm eff}t}|e,G\rangle|^2.
\end{eqnarray}
It is then equivalent to solve the wave function $|\psi(t)\rangle=\alpha(t) |e,G\rangle+\sum_k\beta_k(t)a_k^{\dagger}|g,G\rangle$ under the initial condition $\alpha(0)=1,\beta_k(0)=0$. Considering the short time evolution, the solution is approximately given by~\cite{zeno4},
\begin{equation}
\alpha(t)\approx1-\int_0^{t}dt'e^{i\Omega_1t'}\Phi(t-t')
\end{equation}
where $\Phi(t-t')=\sum_k |g_k|^2e^{-i\omega_k(t-t')}$.

Performing the instantaneous ideal measurements with intervals $\tau$, the survival probability after $m$ measurements is~\cite{zeno4,AQpra10}
\begin{eqnarray}
P_e\left( {t = n\tau } \right)=|\alpha(\tau)|^{2m}=e^{-R(\tau)t},
\end{eqnarray}
where
\begin{equation}
R(\tau)=2\pi \int_{-\infty }^\infty {d\omega F\left({\omega ,{\Omega _1},\tau} \right)} G\left( \omega  \right)
\end{equation}
with
\begin{eqnarray}
F(\omega,\Omega_1,\tau)&=&\frac{\tau}{2\pi}{\rm sinc}^2\left[\frac{(\omega-\Omega_1)\tau}{2}\right],\\
G(\omega)&=&\sum_k |g_k|^2\delta(\omega-\omega_k).
\end{eqnarray}
In our system, the decay rate $R(\tau)$ can be further obtained as
\begin{eqnarray}
R(\tau)&=&\tau \int_{-\infty}^{\infty}d\omega{\rm sinc}^2\left[\frac{(\omega-\Omega_1)\tau}{2}\right]\sum_k|g_k|^2
\delta(\omega-\omega_k)\nonumber\\
&=&\tau\sum_k|g_k|^2
{\rm sinc}^2\left[\frac{(\omega_k-\Omega_1)\tau}{2}\right]\nonumber\\
&=&\frac{g^2\tau}{\pi}\int_{-\pi}^{\pi}dk\{\left[1+\cos(kn)\right]\nonumber\\&&\times{\rm sinc}^2\left[\frac{(\omega-\Omega_1-2\xi\cos k)\tau}{2}\right]\}.
\label{Rtau}
\end{eqnarray}
Under the RWA, we can obtain the similar result, the only difference is to replace $\Omega_1$ in last line of Eq.~(\ref{Rtau}) by $\Omega$.
\begin{figure}
  \centering
  \includegraphics[width=0.5\textwidth]{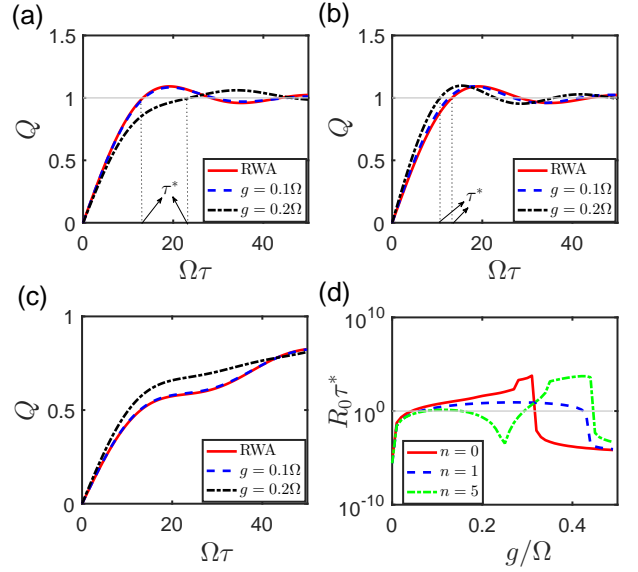}\\
  \caption{(a) (b) (c) $Q(\tau)$ for $n=0,1,4$ respectively. (d) $R_0\tau^*$ as a funciton of coupling strength $g$. The parameters are set as  $\omega_{c}=\Omega$,$\xi=0.1\Omega$.}
  \label{Rtauf}
  \end{figure}
To investigate the QZE and QAZE in our system, we refer to compare $R(\tau)$ and $R_0=R(\infty)$, where the later means the decay rate with an infinite measurement interval, that is, without measurement. In Fig.~\ref{Rtauf}(a,b,c), we plot $Q(\tau)=R(\tau)/R_0$ as a function of the measurement interval $\tau$ for different coupling strength $g$. We note that, under the RWA, the ratio $Q$ is independent of $g$, and is given by the solid curves in Fig.~\ref{Rtauf}(a,b,c). In the weak coupling strength, for example $g=0.1\Omega$, the result based on unitary transformation coincides with that for RWA, which implies the effect of the counter-rotating wave terms can be safely neglected. However, as the increase of coupling strength, the results for unitary transformation derives from that for RWA. For $n=0$ and $n=1$, the results in Fig.~\ref{Rtauf}(a,d) show that the value of $Q$ can be either larger or smaller than one, depending on the measurement interval $\tau$ as well as the coupling strength $g$. For Fig.~\ref{Rtauf}(c), the giant atom with $n=4$, for the considered regime of $\tau$, $Q$ is always smaller than $1$, which implies that QZE occurs. Here, the issue is whether $Q>1$  implies a QAZE.

We recall that the characteristic time of the giant atom is depicted by $1/R_0$ and $\tau$ is the measurement interval, so that if the QAZE occurs, we have to achieve the condition $R_0\tau^*<1$ where $\tau^*$ satisfies $Q(\tau^*)=1$. {Therefore, in Fig.~~\ref{Rtauf}(d), we plot $R_0\tau^*$ as a function of $g$ for $n=0,1,5$ in ultrastrong coupling regime, respectively (the curve of $Q$ as a function of $\tau$ for $n=5$ possesses a similar shape with that of $n=1$, and is not shown here).  For $n=0$ and $n=1$, although one can find $Q>1$ in the some regime with $\tau>\tau^*$, the QAZE does not occur because $R_0\tau^*>1$ for $g=0.2\Omega$, as shown by the red solid and blue dashed curves in Fig.~\ref{Rtauf} (d). On the contrary, we find that the QAZE will occur in the system with $n=5$, in which $R_0\tau^*<1$ in the situation of $g=0.2\Omega$ as shown by the green dot dash line in Fig.~\ref{Rtauf}(d).
The above results show that, the size of the giant atom serves as a degree of freedom, to control how to realize the QZE-QAZE transition by tuning the atom-waveguide coupling strength.

At last, we pay special attention on the case of $n=2$ and $n=6$. Within the RWA, the giant atom will not decay without measurement, that is, $R_0=0$ when $\Omega=\omega_c$~\cite{HY2021}. However, this is not the case in the ultrastrong coupling regime.  In Fig.~\ref{n2}, we plot $R(\tau)$ as a function of $\tau$.  For the coupling strength $g=0.05\Omega$,  $R_0=0$ which also coincides with the RWA, and $R$ will acquire only a small value for both of $n=2$ and $n=6$. As for a relatively strong coupling $g=0.2\Omega$, $R_0$ becomes non-zero due to the atomic frequency shift, which is induced by the counter-rotating wave terms. Performing the measurements, we find that the system undergoes the transition from QZE to QAZE at $\tau\approx1/\Omega$ for $n=2$ in Fig.~\ref{n2}(a) in which $R_0\tau^*<1$. However, for $n=6$,
the fact $R_0\tau^*>1$ as shown in Fig.~\ref{n2}(b) implies that the QAZE will not occur although the value of $Q$ can be larger than $1$.

 \begin{figure}
  \centering
  \includegraphics[width=8.5cm]{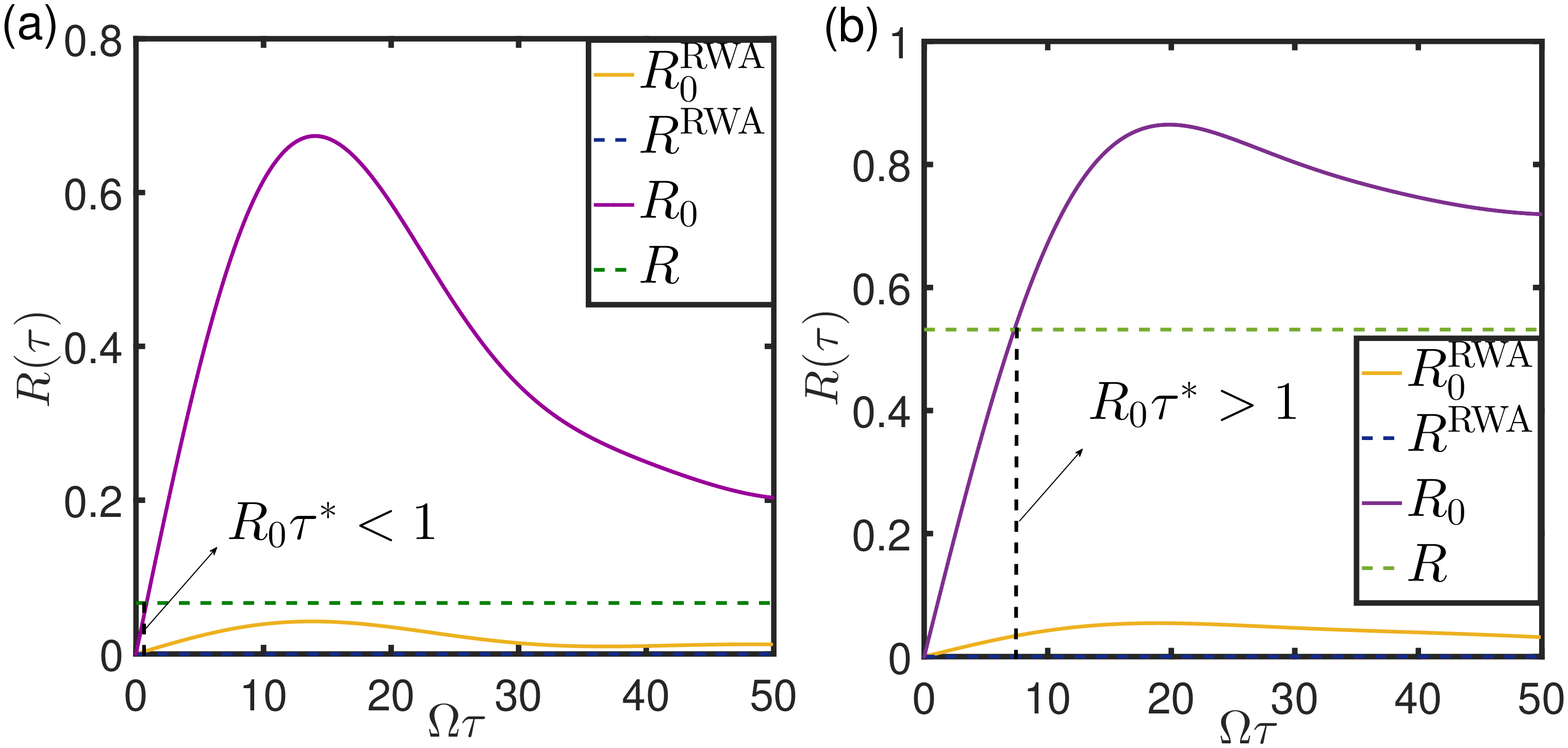}\nonumber\\
  \caption{Decay rate of the giant atom for (a) $n=2$ and (b) $n=6$. {The parameters are set as $\omega_{c}=\Omega$,$\xi=0.1\Omega$, $g=0.2\Omega$ for green and yellow line, $g=0.05\Omega$ for blue and red line.}}
 \label{n2}
\end{figure}

  \section{conclusion}
  \label{conclusion}

In summary, we construct a waveguide QED system by coupling a superconducting circuit to a transmission line and study the controlled QZE and QAZE by tuning the distance of the coupling points and measurement interval. In our model, the superconducting circuit plays as a giant atom
and the transmission line serves as a structured waveguide environment. The ultrastrong coupling between them induces a Lamb shift for the giant atom, which in turn modifies its dissipation process. By introducing the instantaneous ideal measurements, we find the QZE and QAZE for different atomic size. More interesting, we clarify the condition in which the QAZE occurs by comparing the measurement interval and the intrinsic life time (without measurement) of the giant atom.

\begin{acknowledgments}
This work is supported by National Key R$\&$D Program of China (No. 2021YFE0193500), the National Natural Science Foundation of China (Grant Nos. 11875011 and 12175150) and the Natural Science Foundation of Guang-dong Province (Grant No. 2019A1515011400).
\end{acknowledgments}

\end{document}